\documentclass[doublecol]{epl2} 
\usepackage{amssymb}
\usepackage{amsmath}

\title{On the rotation of ONC stars in the Tsallis formalism context}
\shorttitle{Rotation of ONC stars in Tsallis formalism} 

\author{B. B. Soares \and J. R. P. Silva}
\shortauthor{Soares \& Silva}

\institute{Departamento de F\'{i}sica, Universidade do Estado do Rio Grande do Norte, Mossor\'{o}, Brazil

}

\abstract{
The theoretical distribution function of the projected rotational velocity is derived in the context
of the Tsallis formalism. The distribution is used to estimate the average $\langle\sin\,i\rangle$ for a stellar
sample from the Orion Nebula Cloud (ONC), producing an excellent result when compared with observational
data. In addition, the value of the parameter $q$ obtained from the distribution of observed rotations reinforces the
idea that there is a relation between this parameter and the age of the cluster.
}

\begin{document}

\maketitle

\section{Introduction}
The stellar rotation is one of the most challenging problem in astrophysics because it is a complex physical
phenomenon. The stellar rotation has its origin in the stellar formation, when the angular momentum of the progenitor
cloud is transferred to the new stars. At this stage, the stars can suffer the influence of
several physical processes, which can also affect the stellar rotation. Such processes include
proper orbital motion of the cloud, innate differences in internal pressure due to the anisotropy of the cloud and 
the action of the local galactic magnetic field on the parent cloud, as well as shock waves from supernova
explosions. The range of processes as discussed makes the initial range of rotation rates
difficult to understand, but when evolution of rotation rates is considered along with stellar evolution, 
processes such as magnetic braking and structural changes within the star (e.g., development of a radiative 
core) are also important. 

The projected rotational velocity, $V\sin{i}$, where $V$ is the true (equatorial) rotational velocity and $i$ is the
angle between the axis of rotation and the line of sight, can be derived from the Doppler broadening of spectral lines of the
star\cite{shaj_stru29,slettb49,huang53}. However, it is usually difficult to know the inclination of the rotation axis of a 
single star, being possible only in special cases, e.g., when the rotational period can be measured (spotted stars). 
Another special case is in the binary systems with short orbital period, in which the orbital motion is synchronized with the 
rotation one and rotational and orbital axes are parallel\cite{tas_tas92}.

Even given the difficulty in determining the equatorial velocity, one can to derive some global properties of the true
rotation from a frequency distribution of $V\sin{i}$. However, this requires some assumptions
about the frequency of a given angle of inclination of the rotational axis for the group of stars. In general it is
assumed that the axes of rotation of the stars are randomly oriented (e.g.\cite{dien48,brown50,bohm52,bernac70}). 
In this context, the relation between the observed and true rotational velocity for a sample of stars is always 
\(\langle V\sin{i} \rangle/\langle V\rangle=\pi/4\) \cite{chand50}, independently of the environment of stellar formation.

The hypothesis of randomly oriented axes was examined in several works and, although it has not been ruled out, is not
always in agreement with the observed velocity distributions (see discussion in\cite{rhode01}, $\S$ 3.1.2; 
\cite{bernac74}, $\S$ VI). In addition,  later studies showed a
tendency for stars of a given spectral type to present rotation correlated with the galactic coordinates as well as with
the galactocentric distance (e.g.\cite{burki77,demedeiros00}). These studies
open up the possibility of considering a non-random distribution for the inclination of the rotational axes, even if
only in special cases. In fact, several cases in which a power-law fits better than the Gaussian function
(e.g.\cite{soares06,carval07,carval08,carval09,carval10,sant10}), show that the main issue may not just 
be deciding which mathematical function can be used but what statistical theory one should consider. 

It is now well established that a considerable number of stochastic phenomena are better described by power-law
distributions. For example, one could mention the solar neutrino problem\cite{kaniadakis96}, 
peculiar velocities of galaxies\cite{lavagno98} and self-gravitating polytropic
systems\cite{plast_plast93}. Fundamentally it could be included in that list of stochastic
phenomena all the systems whose entropies are more appropriately addressed to $q$-entropy\cite{tsallis88},
postulated as
 \begin{equation}
   S_{q} \equiv \frac{k}{q-1} \left( 1 - \sum_{i=1}^{W} p_i^q \right),
   \label{qsum}
 \end{equation}
where $p_{i}$ denotes the probabilities of the microscopic (individual) configurations. In the limit $q=1$, the
$q$-entropy recovers the standard Boltzmann form. That statistic is therefore a generalization of the
Boltzmann statistics. Usually named nonextensive statistics, the Tsallis thermostatistics states that the entropy
of the sum of two systems is not simply the sum of individual entropies, but provides for cases where such a sum may
result in a greater or lesser entropy. 

The main point of the considerations presented so far is that, besides the
exponential distribution considered in study as Chandrasekhar \& M\"unch\cite{chand50} and Deutsch\cite{deut70}, 
we are often faced with stable
power-law distributions (the $q$-exponentials in equation \ref{q-exp}), since such distributions are ubiquitous in
physical phenomena. The ubiquity is due to the fact that such distributions naturally obey the generalized central limit
theorem (cf.\cite{tsallis95}). Thus, Gaussian-type diffusion and anomalous diffusion, for example, may be
unified into a single scenario which is the nonextensive statistics framework\cite{curado91,tsallis98,prato99,tsallis02}. 

In this paper we present a new way to derive the distribution function of the observed stellar rotations in
the context of the nonextensive statistics. The distribution function is tested with a sample of stellar rotation
from the ONC and accurately reproduces the observational results.
The paper is organized as follows. 
In second section we derive the power-law distribution proposed for this study and the general equations of its moments.
In third section we present the work sample and the method to determine the average
$\langle\sin{i}\rangle$ from the observational sample, and discuss the results.
Fourth section summarizes the main results and in the appendix some relevant calculations is presented.

\section{The power-law distribution function}

The rotational kinetic energy of a star with mass $M$ and radius $R$ rotating as a rigid body around its own axis is
$E_{\rm{rot}}=(1/2)I\omega^{2}$, where $I=(2/5)MR^2$ is the moment of inertia and $\omega=VR^{-1}$ is the angular velocity of the
rotating star. Defining $\omega_{\rm{min}} = ( V\sin{i} )R^{-1}$, it follows that the minimal rotational energy,
$E_{\rm{min}}$ of the star is given by 
 \begin{equation}
   E_{\rm{min}}=(1/5)I\omega_{\rm{min}}^2.
   \label{e_min}
 \end{equation}
Consider a group of stars with similar mass, radius and age, such that the rotational energy $E_{\rm{min}}$ depends only
on $ V\sin{i} $. Suppose further that the energy $E_{\rm{min}}$ is distributed according to the function
$\varphi(E_{\rm{min}})$. The
number of stars with energy $E_{\rm{min}}$ between $ E_{\rm{min}} $ and $E_{\rm{min}} + \,\upd E_{\rm{min}}$ is
 \begin{equation}
   \upd N(E_{\rm{min}})=p(E_{\rm{min}})\,\upd E_{\rm{min}},
   \label{dofunc}
 \end{equation}
where $p(E_{\rm{min}})$ is the probability density of stars with rotational energy $E_{\rm{min}}$.

It is well known from observational studies, that the probability density of observed rotational velocities vanishes with
$ V\sin{i} $. Then it is reasonable to take the distribution function proposed by Deutsch\cite{deut67} as an inspiration
and propose that the probability density decrease according to a generalized exponential law. We are interested in
deriving the distribution of the rotational velocities in the context of the Tsallis formalism. We therefore propose that the
probability density is governed by the non-linear differential equation, $\upd p/\upd x=-\beta p^q$, whose solution is given
by
 \begin{equation}
   p(x)=[1-(1-q)\beta x]^{1/(1-q)}.
   \label{q-exp}
 \end{equation}
This function is the $q$-deformation of the usual exponential function $\exp(-\beta x)$ and reduces to it in the
limit $q\rightarrow 1$ (cf.\cite{borges98}). Accordingly, we can rewrite Equation (\ref{dofunc}) as 
\begin{equation}
\upd N(E_{\rm{min}}) =[1-(1-q)\beta E_{\rm{min}}]^{1/(1-q)}\,\upd E_{\rm{min}}
\end{equation}

from which it follows that
 \begin{equation}
   \upd N(y) =\frac{2}{5}MRy\left[1-(1-q)\frac{y^2}{\sigma^2_y}\right]^{1/(1-q)}\,\upd y,
   \label{dif_y}
 \end{equation}
where $y= V\sin{i} $ and $\sigma_y=(\beta M/5)^{-1/2}$ is the half-width of the distribution.

Equation (\ref{dif_y}) gives the number of stars with a given observed rotational velocity $ V\sin{i} $ from a sample of
stars with similar mass, radius and age. It follows, in the formalism proposed by Tsallis, that the distribution
function of the observed rotation for that group of stars is given, apart from a normalization constant, by
 \begin{equation}
   \varphi_{q}(y)=y\left[1-(1-q)\frac{y^{2}}{\sigma_{y}^{2}}\right]^{1/(1-q)}.
   \label{fqobv}
 \end{equation}
The above equation has also been derived from a different way by Soares\etal\cite{soares06}.

In accordance with Kraft\cite{kraft65}, we can write the distribution function of true rotations, apart from a
normalization constant, as 
 \begin{equation}
  F_{q}( V) =
   V^{2}\left[1-(1-q)\frac{V^{2}}{\sigma_{V}^{2}}\right]^{1/(1-q)}
  \label{fqtru}
 \end{equation}
where $\sigma_{V}$ is the half-width of the distribution. Since the function (\ref{q-exp}) recovers the usual
exponential in the limit $q\rightarrow 1$, Equations (\ref{fqobv}) and (\ref{fqtru}) recover the standard
form of $\varphi(y)$ and $F(V)$\cite{deut70,kraft65}, respectively, when $q\rightarrow 1$.

\subsection{Moments of the distributions}

The $r$-th moment of the distribution $F_{q}( V)$ is calculated by making 
 \begin{equation}
   \langle V^{r}\rangle = \int_{0}^{ V_{{\rm{max}}}}\, V^{r}\,F_{q}( V)\,\upd V,
   \label{momF1}
 \end{equation}
where \(V_{\rm{max}}=\sigma_{V}/\sqrt{1-q}\) for $q<1$, or \(V_{\rm{max}}=\infty\) for $q\ge 1$, 
in order to ensure the positivity of the function. 
Hence Eq. (\ref{momF1}) provides 
 \begin{equation}
   \langle V^{r}\rangle= \frac{2}{\sqrt{\pi}}\left(\frac{\sigma_{V}}{\sqrt{1-q}}\right)^{r}
   \frac{\Gamma\left(\frac{r+3}{2}\right)\Gamma\left(\frac{7-5q}{2-2q}\right)}
   {\Gamma\left(\frac{2-q}{1-q}+\frac{r+3}{2}\right)}
   \label{momF2}
 \end{equation}
for $q<1$, and
 \begin{equation}
   \langle V^{r}\rangle = \frac{2}{\sqrt{\pi}}\left(\frac{\sigma_{V}}
   {\sqrt{q-1}} \right)^{r}\frac{\Gamma\left(\frac{r+3}{2}\right)
   \Gamma\left(\frac{1}{q-1}-\frac{r+3}{2}\right)}{\Gamma\left(\frac{5-3q}
   {2q-2}\right)}
   \label{momF3}
 \end{equation}
for $q\ge 1$. The moments of order $r=1$, $r=2$ and $r=3$ in Equations (\ref{momF2}) and (\ref{momF3}) give the
mean, mean-square deviation, and asymmetry of $F_{q}( V)$, respectively. In these cases, however, it
should be noticed that the convergence is achieved only up to certain values ​​of $q$: $q<3/2$ for $r=1$, $q<7/5$ for
$r=2$ and $q<4/3$ for $r=3$. 

Similarly to the previous case, one can determine the moments of $\varphi_{q}(y)$ as being 
 \begin{equation}
   \langle y^{r}\rangle = \left(\frac{\sigma_{y}}{\sqrt{1-q}}\right)^{r}
   \frac{\Gamma\left(\frac{r+2}{2}\right)\Gamma\left(\frac{3-2q}{1-q}\right)}
   {\Gamma\left(\frac{2-q}{1-q}+\frac{r+2}{2}\right)},
   \label{momF5}
 \end{equation}
and
 \begin{equation}
   \langle y^{r}\rangle = \left(\frac{\sigma_{y}}{\sqrt{q-1}} \right)^{r}
   \frac{\Gamma\left(\frac{r+2}{2}\right)\Gamma\left(\frac{1}{q-1}-\frac{r+2}{2}
   \right)}{\Gamma\left(\frac{2-q}{q-1}\right)}.
   \label{momF6}
 \end{equation}
for $ q < 1 $ and $ q \ge 1$ respectively.
As in the previous case, the moments of order $r=1$, $r=2$ and $r=3$ in Equations (\ref{momF5}) and
(\ref{momF6}) give the mean, mean-square deviation, and asymmetry of $\varphi_{q}(y)$, respectively. For
the regime $q\ge 1$, the convergence of the moments for $y$ is: $q<5/3$ for $r=1$, $q<3/2$ for $r=2$, and $q<7/5$ for
$r=3$.

\subsection{Relation between the moments}

Chandrasekhar \& M\"unch\cite{chand50} had shown that when $y=x\sin{i}$, with
$\varphi(y)$ being affected by the randomness factor $\sin{i}$, there is a relation between the moments of these frequency
functions which enables us to pass from the moments of the one to the moments of other. As is pointed out in the same
work, the relation is still valid, even when $y = x\psi(i)$, where $\psi(i)$ is an arbitrary integrable function.
Using Eqs. (\ref{momF2}), (\ref{momF3}), (\ref{momF5}) and (\ref{momF6}), we have obtained a similar relation between the moments of the distribution of $\varphi_{q}(y)$
and $F_{q}( V)$:
\begin{equation}
\frac{\langle y^{r}\rangle}{\langle V^{r}\rangle} = 
\frac{\sqrt{\pi}}{2}\left(\frac{\sigma_{y}}{\sigma_{V}}\right)^{r}
\left[\frac{\Gamma\left(\frac{r+2}{2}\right)
\Gamma\left( \frac{3-2q}{1-q}\right)
\Gamma\left(\frac{2-q}{1-q}+\frac{r+3}{2}\right)}
{\Gamma\left(\frac{r+3}{2}\right)
\Gamma\left(\frac{7-5q}{2-2q}\right)
\Gamma\left(\frac{2-q}{1-q}+\frac{r+2}{2}\right)}\right]
\label{qrmom1}
\end{equation}
for $q<1$, and 
\begin{equation}
\frac{\langle y^{r}\rangle}{\langle V^{r}\rangle} = 
\frac{\sqrt{\pi}}{2}\left(\frac{\sigma_{y}}{\sigma_{V}}\right)^{r}
\left[\frac{\Gamma\left(\frac{r+2}{2}\right)
\Gamma\left( \frac{5-3q}{2q-2}\right)
\Gamma\left(\frac{1}{q-1}-\frac{r+2}{2}\right)}
{\Gamma\left(\frac{r+3}{2}\right)
\Gamma\left(\frac{2-q}{q-1}\right)
\Gamma\left(\frac{1}{q-1}-\frac{r+3}{2}\right)}\right]
 \label{qrmom2}
\end{equation}
for $q\ge 1$. The order $r=1$, $r=2$ and $r=3$ in Equations (\ref{qrmom1})
and (\ref{qrmom2}) give the relations between the mean, mean-square deviation, and asymmetry of the
distribution functions $F_{q}( V)$ and $\varphi_{q}(y)$. It is important to notice that in the limit $q\rightarrow 1$ all
$q$-dependent terms in Equation (\ref{qrmom2}) trend to $\Gamma\left(\frac{1}{q-1}\right)$ and therefore
 \begin{equation}
   \lim_{q\to 1}\frac{\langle y^{r}\rangle}{\langle V^{r}\rangle}= 
   \frac{\sqrt{\pi}}{2}\frac{\Gamma\left(\frac{r}{2}+1\right)}{\Gamma\left(\frac{r}{2}+\frac{3}{2}\right)},
   \label{cmrmom}
 \end{equation}
where we have assumed $\sigma_{y}=\sigma_{V}$. Equation (\ref{cmrmom}) is precisely the relation between moments found by
Chandrasekhar \& M\"unch (\cite{chand50}, Eq. 18).

\vspace{0.2cm}
The proposed general power-law function (\ref{fqobv}) can describe distributions of projected rotational velocities
for stars from different populations, depending on the value of the parameter $q$. As we will discuss 
later, the parameter $q$ constitutes a link between the theoretical distribution and the observational one.

\section{Confronting model with observational data}

\subsection{The sample}

Our sample was selected out of the data provided by Rhode\etal\cite{rhode01} to match the following criteria: projected rotation, $V\sin{i}$, and true rotation, $V$, greater than $\sim 11$\,km/s; and $\sin\,i\leq 1$. Applying these criteria
we obtained a sample of 86 stars whose $\sin\,i$ parameters are accurately determined. Stellar radii are
estimated from combining the luminosities, $L$, and effective temperatures, $T_{eff}$, displayed in
\cite{rhode01} (Table 1), according to the relation $R\propto L^{1/2}T^{-2}$. The stellar parameters $L$ and
$T_{eff}$ are uncertain by about 0.2 and 0.02\,dex, respectively (see Hillenbrand\cite{hillenb97}). Adding these
uncertainties in quadrature, we have that the error in $R$ is less than 0.11\,dex. The masses and spectral types for
stars in our sample can be found in \cite{hillenb97}.

The main features of the stars in our sample are: (i) age less than 1--2\,Myr; (ii) spectral types ranging
from G6 to M5, with 89\% of that stars ranging from K0 to M4; (iii) stellar masses covering a range between
0.1\,M$_{\odot}$ and 2.7\,M$_{\odot}$, with median 0.3\,M$_{\odot}$; and (iv) stellar radius, $R$, ranging from about
1.2\,R$_{\odot}$ to 8.2\,R$_{\odot}$, with a median of 2.4R$_{\odot}$. Then our sample consists of stars
whose main parameters are very similar, as required for this study.

The projected rotational velocities were obtained from the high-dispersion spectra of stars by using the
Fourier cross-correlation method. The smallest $ V\sin{i} $ value that is taken to be reliably measured is 11.0\,km/s. 
For a complete discussion on the observational procedure and error analysis of $V\sin{i}$ data, the reader is
referred to \cite{rhode01}. The rotations have been calculated from measurements of the stellar
rotational period, $P$, and radius, $R$, using the equation $V=2\pi RP^{-1}$. The calculated rotation values
​​have an uncertainty around 11\%. The main source of uncertainty in these values are the stellar radii, as
measured periods have a typical accuracy of 1\% or better (cf.\cite{rhode01}, see also Choi \& Herbest
\cite{choiherb96}).

\subsection{The average $\langle\sin{i}\rangle$}

\begin{figure}[ht]
   \centering
   \includegraphics[width=8.5cm]{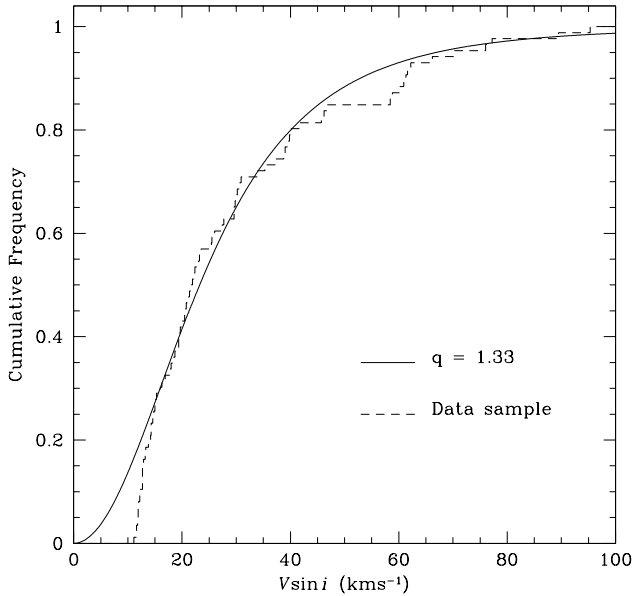}
      \caption{Cumulative distribution of the observed rotational data, $ V\sin{i} $, and the integral curve of the proposed
      function that best fits the distribution. The best-fit parameters are $q=1.33\pm
      0.03$ and $\sigma=20.9\pm 0.88$.}
      \label{acum_dist}
  \end{figure}

\begin{figure}[ht]
   \centering
   \includegraphics[width=8.5cm]{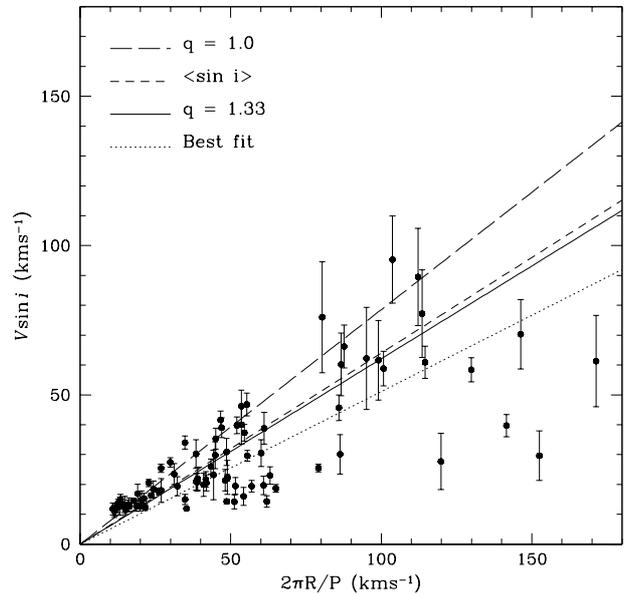}
   \caption{Observed rotation, $ V\sin{i} $, as a function of rotation, $V=2\pi RP^{-1}$, for the 86 stars of the sample.
    The curve $q = 1$ represents the function $\langle V\sin{i} \rangle/\langle V\rangle = \pi/4$; $\langle \sin{i}
    \rangle$ indicates $\langle V\sin{i} \rangle/\langle V\rangle =0.64$; q = 1.33 marks the position of the curve
    $\langle V\sin{i} \rangle/\langle V\rangle =0.62$; and the last one shows the best-fit curve 
    $\langle V\sin{i} \rangle/\langle V\rangle =0.51$, included here only for comparison with the others. The error bars on
    $ V\sin{i} $ are from \cite{rhode01}.}
   \label{v_vsini}
 \end{figure}

In this section we show that Function (\ref{fqobv}) reproduces accurately the average inclination
angles of the rotational axes, $\langle\sin{i} \rangle$, for stars in the sample. First we have determined the
values for the parameters $q$ and $\sigma_y$ of the curve that best fits the distribution of the projected
rotational velocities. In order
to avoid biases due to the choice of the bin when using histograms to represent frequency distributions, we fit the
integral of Equation (\ref{fqobv}) to the cumulative distribution of the projected
rotational velocities. Figure \ref{acum_dist}
shows the cumulative distribution and the integral of Equation (\ref{fqobv}) that best fits the distribution. We have
found $q=1.33\pm 0.03$ and $\sigma_y=20.9\pm 0.88$. To adjust the curve to the distribution of $V\sin{i}$
we have used a
Levenberg-Marquardt nonlinear least squares algorithm. The fit parameters are $\chi^2/dof \approx 2.01\times 10^{-3}$,
for a number of degrees of freedom $dof=84$, and probability $P>99$\%. The region of low rotations, by reason of lack of
measurements of $V\sin{i}$, presents the largest differences between observational data and the model.
This lack of
the data is due to the impossibility for detecting precise rotational velocities below 11\,km/s in the observational
campaign of Rhode\etal\cite{rhode01}.
The observed average sine of the inclination of the rotational axes of the ONC stars is $\langle\sin{i}\rangle =
0.64\pm0.07$. This average was calculated by combining the values ​​of $V\sin{i}$ and $V$ for
our sample. The uncertainty in this value is mainly due to the error of 11\%, associated with the rotation, $V$.
Assuming a random
distribution of stellar rotational axes, the mean value is expected to be $\pi/4=0.79$, which is very different from the
observed value. This is also displayed in Figure \ref{v_vsini}, where we represent the distribution of $ V\sin{i} $ as a
function of $V$. In this figure, the majority of points are well distributed around the curve $\langle V\sin{i}
\rangle/\langle V\rangle =0.64$, but the curve $\langle V\sin{i} \rangle/\langle V\rangle = \pi/4$ is located near the
upper limit of the set of points. Rhode\etal\cite{rhode01} discuss this discrepancy between the observed average of
$\sin{i}$ and theoretical expected value. They considered that a possible explanation is that the rotational axes in
young clusters such as ONC could be aligned rather than oriented randomly. In fact, the stars in association are formed
from a cloud of interstellar material in a relatively short interval of time. In the formation process, the angular
momentum of the parent interstellar cloud is transferred to the newly-formed stars. It is therefore reasonable to assume
that the orientation of the angular momentum of these stars initially reflect the momentum of the rotating progenitor
cloud. Thus, a association of very young stars tends to have a preferential orientation of axes of rotation, although
this feature can be lost gradually over time (cf.\cite{vveer75,struve50}).

By using Equation (\ref{qrmom2}) with parameter $q=1.33$, as obtained from the fit of the $ V\sin{i} $ cumulative
distribution, we calculate the expected value for the average $\langle\sin{i}\rangle=0.62\pm0.04$. This value is
approximately equal to the average of $\sin{i}$ observed, as shown in Figure \ref{v_vsini}, where the curve
$\langle V\sin{i} \rangle/\langle V\rangle =0.62$ almost coincides with the observed one $\langle V\sin{i} \rangle/\langle
V\rangle =0.64$. The expected average, calculated in accordance with our model, has a connection with the observational
data through the parameter $q$, which is derived from the distribution of the projected
rotational velocities. It is therefore understandable that our
proposal provides an expected average value of $\langle\sin{i}\rangle$ close to the observed one. Our model is
particularly important in cases where only the projected rotational velocities are
available, because it can provide an average $\langle\sin{i}\rangle$ closer the true observed value than can the
standard model. The standard relation
$\langle V\sin{i} \rangle/\langle V\rangle=\pi/4$ is appropriate only in cases where there is no preferential
orientation of stellar angular momenta. The relation expressed in Equation (\ref{qrmom2}) can be used in any
situation, even if the distribution is completely random ($q=1$). 

In line with the discussion above, if one assumes that the younger the cluster the greater the tendency for a
preferential orientation of stellar angular momenta, the parameter $q$ can be associated
with the degree of randomness of the angular momentum orientations. It can also be related to the age of the cluster.
Using Equation (\ref{fqobv}), Soares\etal\cite{soares06} have analyzed the distribution of a sample of
projected rotational velocities of stars in the Pleiades cluster (115\,Myr) and have found $q=1.38$.
Utilizing the same method, 
Santoro\cite{sant10} has found $q=1.51$ for the Hyades cluster (665\,Myr). These results, in combination with the
present
study, reinforce the idea that the parameter $q$ is related to the age of the clusters. However, a detailed study with
stellar rotational data from clusters of different ages using the methodology presented here in this work is needed in
order to determine whether such a relation exists. In order to confirm this relation, it is necessary to verify that
$q=1$ for the oldest clusters.

\section{Conclusions}

We have derived the distribution function of the projected rotational velocity in the context of the Tsallis formalism. The
distribution function is $q$-dependent and recovers the Deutsch's function in the limit $q\rightarrow 1$. The parameter
$q=1.33\pm 0.03$ was determined from a sample of 86 stellar projected rotational velocities from
the ONC. The expect average $\langle\sin{i}
\rangle=0.62\pm0.04$ was calculated using the relation between the first moments of the theoretical distribution
functions of the observed, $V\sin{i}$, and the true, $V$, rotational
velocities in the Tsallis formalism. The result reproduces accurately the average
$\langle\sin{i} \rangle = 0.64\pm 0.07$ from the observational data. The procedure presented in this work
constitutes an
efficient method to determine the relation between the moments of the distributions of the observed and the true stellar
rotational velocities. We suggest and discuss the possible existence of a relation between the parameter $q$ and the degree of
randomness of the angular momentum orientations of stars. As well we suggest the possible existence of a relation between the parameter $q$ and the age of the
stellar clusters. However, a detailed study involving clusters of different ages is needed in order to determine
whether any relation exists. This study is being developed for a forthcoming paper.

\acknowledgments
This study was partially funded by the Programa Institutos Nacionais de Ci\^encia e Tecnologia
(MCT-CNPq-Edital N$^o$ 015/2008). We should like to thank Dr. Thomas Dumelow for critically reading the manuscript.

\end{document}